# MEDICAL IMAGE HARMONIZATION USING DEEP LEARNING BASED CANONICAL MAPPING: TOWARD ROBUST AND GENERALIZABLE LEARNING IN IMAGING

Vishnu M. Bashyam[1*], Jimit Doshi[1], Guray Erus[1], Dhivya Srinivasan[1], Ahmed Abdulkadir[1], Mohamad Habes[3], Yong Fan[1], Colin L. Masters[4], Paul Maruff[4], Chuanjun Zhuo[5,6], Henry Völzke[8,19], Sterling C. Johnson[9], Jurgen Fripp[10], Nikolaos Koutsouleris[11], Theodore D. Satterthwaite[1,12], Daniel H. Wolf[12], Raquel E. Gur[7,12], Ruben C. Gur[7,12], John C. Morris[13], Marilyn S. Albert[14], Hans J. Grabe[15], Susan M. Resnick[16], R. Nick Bryan[17], David A. Wolk[2], Haochang Shou[18], Ilya M. Nasrallah[7] and Christos Davatzikos[1*], for the iSTAGING and PHENOM consortia

[1] Artificial Intelligence in Biomedical Imaging Lab, University of Pennsylvania, Philadelphia, PA, USA
[2] Department of Neurology, University of Pennsylvania
[3] Biggs Alzheimer's Institute, University of Texas San Antonio Health Science Center, USA
[4] Florey Institute of Neuroscience and Mental Health, University of Melbourne
[5] Tianjin Mental Health Center, Nankai University Affiliated Tianjin Anding Hospital, Tianjin, China
[6] Department of Psychiatry, Tianjin Medical University, Tianjin, China
[7] Department of Radiology, University of Pennsylvania
[8] Institute for Community Medicine, University of Greifswald, Germany
[9] Wisconsin Alzheimer's Institute, University of Wisconsin School of Medicine and Public Health
[10] CSIRO Health and Biosecurity, Australian e-Health Research Centre CSIRO
[11] Department of Psychiatry and Psychotherapy, Ludwig Maximilian University of Munich
[12] Department of Psychiatry, University of Pennsylvania
[13] Department of Neurology, Washington University in St. Louis
[14] Department of Neurology, Johns Hopkins University School of Medicine
[15] Department of Psychiatry and Psychotherapy, Ernst-Moritz-Arndt University
[16] Laboratory of Behavioral Neuroscience, National Institute on Aging
[17] Department of Diagnostic Medicine, University of Texas at Austin
[18] Department of Biostatistics, Epidemiology and Informatics, University of Pennsylvania
[19] German Centre for Cardiovascular Research, Partner Sit Greifswald, Germany

## ABSTRACT

Conventional and deep learning-based methods have shown great potential in the medical imaging domain, as means for deriving diagnostic, prognostic and predictive biomarkers, and by contributing to precision medicine. However, these methods have yet to see widespread clinical adoption, in part due to limited generalization performance across various imaging devices, acquisition protocols and patient populations. In this work, we propose a new paradigm in which data from a diverse range of acquisition conditions are "harmonized" to a common reference domain, where accurate model learning and prediction can take place. By learning an unsupervised image to image canonical mapping from diverse datasets to a reference domain using generative deep learning models, we aim to reduce confounding data variation while preserving semantic information, thereby rendering the learning task easier in the reference domain. We test this approach on two example problems, namely MRI-based brain age prediction and classification of schizophrenia, leveraging pooled cohorts of neuroimaging MRI data spanning 9 sites and 9701 subjects. Our results indicate a substantial improvement in these tasks in out-of-sample data, even when training is restricted to a single site.

------------------------------

[*] Corresponding authors: Vishnu Bashyam and Christos Davatzikos, {Vishnu.Bashyam; Christos.Davatzikos}@pennmedicine.upenn.edu, 3700 Hamilton Walk, 7th Floor, Artificial Intelligence in Biomedical Imaging Lab, University of Pennsylvania, Philadelphia, PA 19104; https://www.med.upenn.edu/cbica/

1. INTRODUCTION

Deep learning models can produce useful results on a variety of prediction tasks in medical imaging, including segmentation [1, 2], precision diagnostics [3, 4], and prediction of clinical outcome [5, 6], however, they often perform inconsistently when applied to data obtained under different conditions such as imaging devices, acquisition protocols, and patient populations [7, 8]. This imaging heterogeneity can diminish the generalizability of prediction models as they also reflect irrelevant or confounding features. In the medical imaging domain, poor generalization presents a major limitation to the widespread clinical adoption of deep learning based predictors.

Restrictions in generalizability is a common characteristic of modeling high dimensional data with many degrees of freedom, though it can be overcome partially if sufficiently large and diverse training sets are available [9]. While there has been progress recently arising from efforts to collect very large and diverse training sets through pooling data across multiple studies [10], it is often the case that the desired labeled data (ie. clinical or genomic variables) is available only for a subset of the studies, whereas unlabeled data is more readily available. Importantly, even if a sufficiently large and diverse labelled dataset can be brought together to support extensive training of a machine learning model at present, continuous advances in imaging protocols as well as in biomarker and clinical measurements will change the characteristics of the data, thereby raising the need for re-acquiring a new training dataset. Therefore, machine learning-based methods have seen limited applicability in the clinic, at least relative to their potential.

Herein, we address the aforementioned challenges by explicitly mapping the neuroimaging data to a canonical reference domain in which the appearance of the images is expected to be invariant to the source of acquisition. This complex non-linear stationary mapping is implemented by a fully convolutional generative adversarial network (GAN) that reduces acquisition-related confounding variability while retaining semantic information. GANs, a family of deep learning methods using adversarial training to learn a generative model of a target distribution, have proven to be highly effective at synthesizing imaging data in a wide range of scenarios [11]. Recently, advances in the computer vision community have shown how GANs can be used to train deep learning models that are robust to adversarial perturbations [12-15]. In the case of modeling natural variation, Robey et al. [16] used learnable models of variation, to train robust deep learning for factors such as weather conditions in street sign recognition and background color in digit recognition. Despite these advances, in the medical imaging community there is still difficulty in dealing with multisite data and generalizing predictors to new sites. However, there have been encouraging results using the GAN approach to model site-based variation in medical images. In particular, Y. Gao et al. [17] used a GAN based approach for intensity normalization on multi-site T2-FLAIR MRI data and Modanwal et al. [18] developed a GAN based image harmonization approach for dynamic contrast-enhanced breast MRI scans. Additionally, some other deep learning based approaches to site harmonization have been proposed, but their usefulness is limited by factors such as requiring training data to include paired subjects whom have been imaged at both sites [19-21]; a condition which is very difficult to meet in practice with sufficiently large and continuously updated training samples.

In this work we employ a modified version of CycleGAN, a subclass of GANs. CycleGAN is able to model domain level variation from unpaired data, while preserving semantic information within the data [22]. This is accomplished by learning a mapping from data in domain A to the reference domain B, while ensuring that transformed data retain their original semantic information by enforcing an identity mapping back to the original domain. In effect, this matches variation of the confounding dimensions to that of the template site.

We evaluate our proposed approach on MRI neuroimaging data, by showing improvements in brain age estimation and schizophrenia classification on individuals imaged at sites other than those of the training data. We leverage multiple neuroimaging datasets spanning 9 sites and 9701 individuals, to learn a GAN model of confounding image variations and evaluate comprehensively out-of-sample prediction performance.

## 2. RESULTS

### 2.1 Capturing confounding inter-scanner variations via CycleGAN

Visually, image characteristics became markedly more uniform in terms of grey/white matter contrast, overall and regional intensity, and noise patterns, after canonical mapping via CycleGAN. Characterized by the intensity distributions of tissue types, shown in figure 1, it can be seen that the regional intensities of the mapped image aligned more closely with that of the image from the reference domain. Additionally, it can be seen that the mean intensity and the standard deviation maps of the mapped image is more similar to the reference domain. The preservation of predictive information will be evaluated in the age prediction and schizophrenia classification experiments.

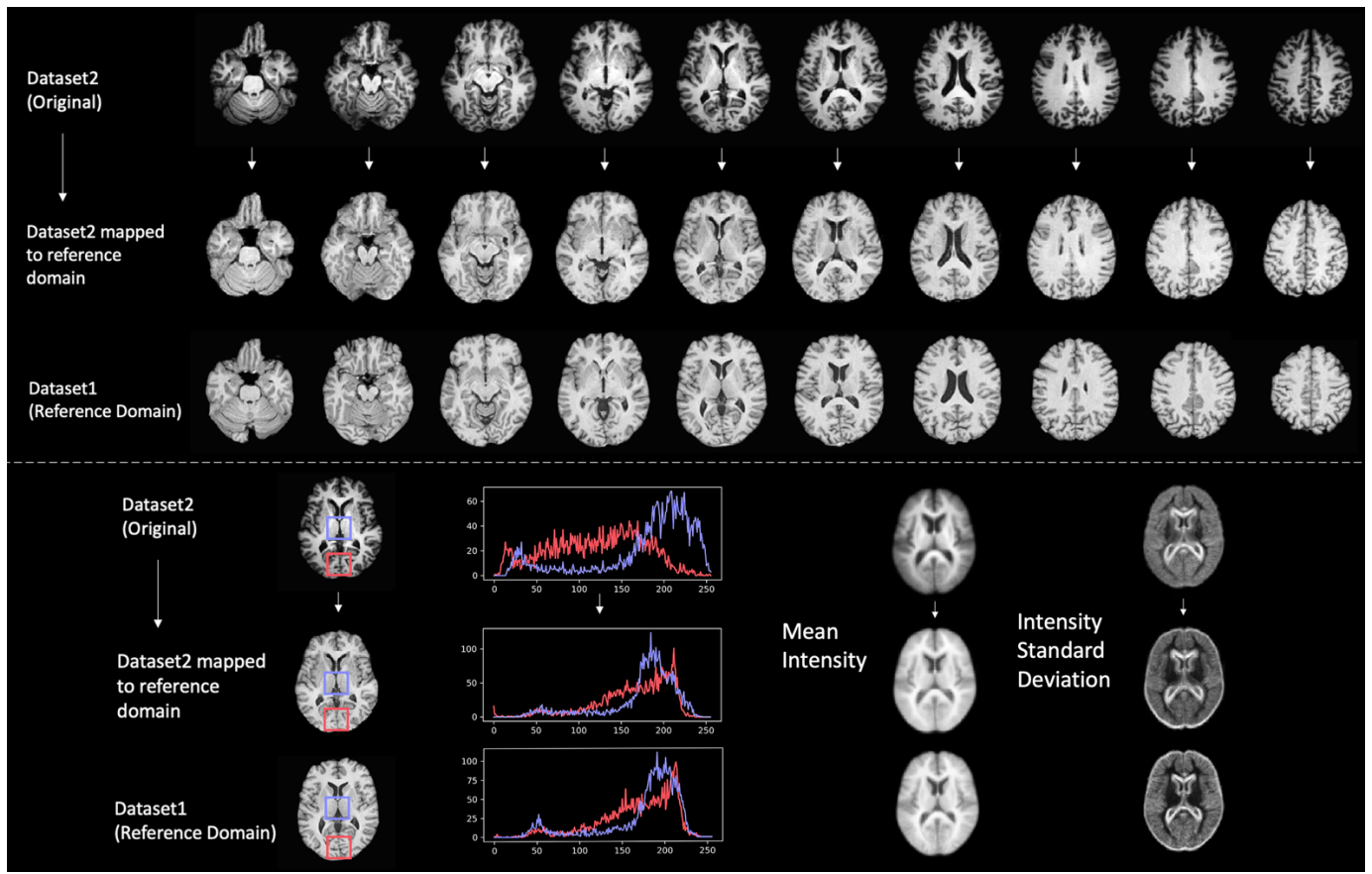

Figure 1: Top: Representation of mapping of axial slices of participant from Dataset2 to the reference domain. Bottom left: Comparison of regional histograms before and after mapping to reference domain. Bottom right: Mean and standard deviation maps across all scans.

### 2.2 Age Prediction

The first evaluation task, namely age prediction, was performed using two large datasets (Dataset1, a study performed on a 1.5T scanner, and Dataset2, a study performed on a 3T scanner) that cover similar age ranges. To establish a baseline, we tested the age prediction performance by training our predictor on data from Dataset2 and testing on data from Dataset1, where we observed poor generalization ability (Mean Absolute Error (MAE) = 15.63 years, Pearson correlation coefficient = 0.304).

Taking Dataset1 as our reference domain, we then trained our CycleGAN model to learn the transformation mapping between Dataset2 and Dataset1. To obtain our harmonized data, we applied the forward mapping of this model to the Dataset2 data. Using the domain adapted data, we saw a large improvement in the generalization performance of our predictor (MAE = 7.55 years, Pearson correlation coefficient = 0.791).

| HARMONIZED | MAE | CORRELATION |
|---|---|---|
| No | 15.63 | 0.304 |
| Yes | 7.55 | 0.791 |

Table 1: Age prediction results with model trained on Dataset1 and tested on Dataset2

Next, we evaluated our harmonization method in the multisite setting. In this experiment, we used our modified CycleGAN network (see supplemental 1 for more information) to map five datasets to the canonical reference domain and evaluated age prediction performance on a network model trained only on the 2739 scans originally from the reference dataset (Dataset1, in this experiment). We observed consistent improvement in age prediction performance across all sites in terms of MAE and Pearson correlation. With an overall improvement in terms of mean absolute error from 9.85 years to 6.68 years and correlation with true age from 0.271 to 0.872.

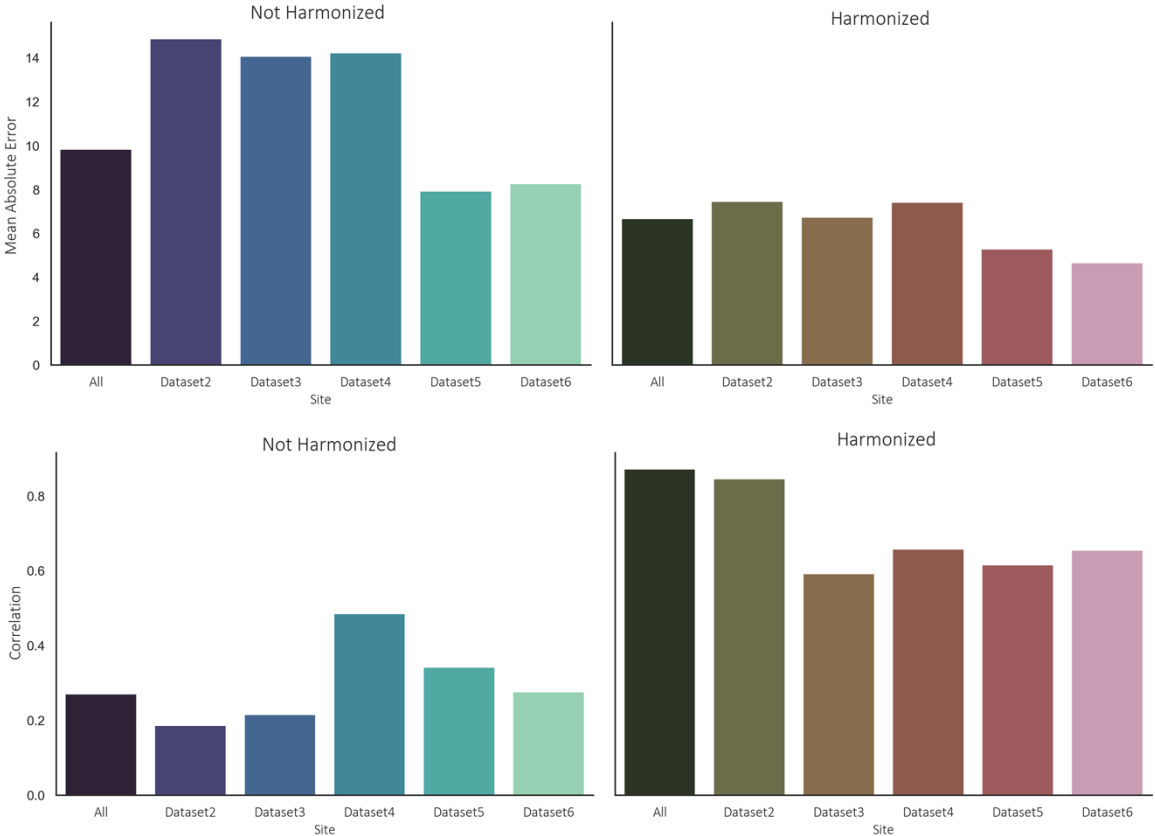

Figure 2: Multisite age prediction results with and without harmonization. Model was only trained with scans from the reference domain (Dataset1)

## 2.3 Schizophrenia Classification

We further evaluate the accuracy of the method to improve site generalization on classification of MRI scans from patients with schizophrenia. While an ample literature on MRI-based classification of schizophrenia exists, prior work was based on extensive and specialized preprocessing of images and extraction of pre-determined features. Such procedures have limited the application of these methods in the clinic, in part due to lack of good generalization to data from new scanners. Our experiments tested the effectiveness of constructing a deep learning classifier directly from raw MRI scans, with and without the use of canonical mapping harmonization. For this we used a dataset containing scans from three separate sites, each with unique imaging characteristics. For these experiments, we harmonized the testing site to the training site. In the case where two sites were used for training, the first site listed was used as the reference site and the other two sites were harmonized to it. We observed a substantial improvement in classification performance after the test datasets are harmonized to the training domain.

| TRAIN/TEST | HARMONIZED | AUC |
|---|---|---|
| SITE 1*/SITE 2 + SITE 3 | No | 0.54 |
| | Yes | 0.71 |
| SITE 2*/SITE 1 + SITE 3 | No | 0.52 |
| | Yes | 0.70 |
| SITE 1* + SITE 2/SITE 3 | No | 0.50 |
| | Yes | 0.76 |
| SITE 1* + SITE 3/SITE 2 | No | 0.56 |
| | Yes | 0.74 |
| SITE 2* + SITE 3/SITE 1 | No | 0.54 |
| | Yes | 0.73 |

*Reference Site

Table 2: Schizophrenia classification results showing improved performance with harmonization to the reference domain

## 3. DISCUSSION

In the computer vision community, work related to improving the generalizability of deep learning based predictors has centered around the use of large and diverse training sets, whereby the predictor can learn high-level feature representations distinct from confounding variation. In the medical imaging context, at present, it is not realistic to collect such varied training sets for each desired task. While large and diverse imaging datasets have become available, subjects in these sets have limited and varying labeled information available (e.g. disease, pathology, or genomic labels). Moreover, as the characteristics of images change with time, one would need to recreate such large labeled datasets, a task that is impractical for most clinical problems. Therefore, we propose a paradigm by which confounding variation will be removed via canonical mapping across scans, thereby enabling smaller less diverse datasets to be useful for model construction and robust to subsequent variations of image characteristics. Critically, the canonical mapping model derived from unlabeled datasets can continue to evolve, as image acquisition itself evolves, hence allowing for new types of data to be canonically mapped, and therefore retaining the value of classification models derived from relatively limited and precious labeled scans. Such a method has the potential to enable the widespread adoption and standardization of deep learning based methods, both because it avoids the need for specialized and sophisticated processing of the images, but also because of the good generalization properties achieved via canonical mapping.

We proposed an image-level harmonization method capable of learning a robust canonical mapping to a reference domain. Our harmonization schema attempts to identify the complex and non-linear imaging variation occurring due to confounding factors, such as scanners, acquisition protocols, and cohorts, while maintaining inter-subject variation related to classification labels. Through harmonization, we observe improvements in image consistency with the reference domain, specifically in terms of grey/white matter contrast, overall intensity, and noise patterns. Through our experiments we found our harmonization method was able to successfully learn this mapping even in a site as small as 90 scans, as long as the reference domain was sufficiently large. While prior work has shown that CycleGAN can be a powerful method for image-level domain adaptation [22], in the medical imaging context the preservation of fine anatomical structure and predictive information is critical. While there is certainly a risk of overfitting and removing non-site variation when using highly non-linear methods, we demonstrate the preservation of this predictive information in neuroimaging data through our experiments on brain age estimation and schizophrenia classification.

Brain age estimation has become an established biomarker of overall brain health in the neuroimaging community, exhibiting overlapping neuroanatomical patterns with a variety of other pathologic processes [23, 24]. Accurate brain age estimation is dependent on fine neuroanatomical patterns that can be obfuscated by imaging variation across site. Therefore, it is a prime candidate to assess harmonization performance.  We demonstrate improved age prediction generalization in five separate sites, following their mapping to the reference domain, in which the predictor was trained.

The popularity and success of deep learning based brain age estimation may partly be attributed to wide availability of age labels across neuroimaging datasets. This has enabled models to learn invariance to the confounding effects of site variation. While this is true of age prediction, it is not the case for schizophrenia classification. The availability of labeled schizophrenia neuroimaging data is relatively sparse and thus robust prediction models have remained limited. Further, perhaps even more so than in age prediction, the neuropathologic signal of schizophrenia is indistinct and varied, with possible structural abnormalities presenting in the prefrontal cortex, basal ganglia, thalamus, hippocampus, and medial temporal lobe [25].  We demonstrate a substantial improvement in the performance of schizophrenia prediction in out-of-sample data, even when the prediction model is trained on a single site with relatively limited data.

We recognize that the prediction performance in age prediction and schizophrenia classification, particularly in the harmonized case, can certainly be improved with more specialized prediction networks, optimization techniques, hyperparameter selection, and larger sample sizes, but we aim to simply demonstrate improvements with harmonized data with an non-optimized, commonly used network and a reasonable sample size, in view of the scarcity of such datasets. We anticipate future work will incorporate harmonization to a reference domain with finely tuned networks to construct powerful and generalizable imaging predictors. Additionally, further work is required to investigate how image level harmonization techniques such as this behave when the reference domain and out of sample domains differ sharply across covariates (such as age, ethnicity, pathology). Future directions of this work could involve the explicit modeling of such covariates directly within the network.

## 4. METHODS

### 4.1 Datasets

### 4.1.1 Age Prediction

For age prediction with a single out of sample site, we use two large datasets of T1-weighted brain MRI scans that cover a wide range of ages, Dataset1 (n = 2739) and Dataset2 (n = 952). See Table 3 for more details.

For multisite age prediction, we use six datasets. Five (n = 6137) are used for the out of sample evaluation and Dataset1 is used as the canonical reference domain and training data from the age prediction model. See Table 3 for more details.

| TOTAL SUBJECTS = 8876 | Dataset1 (REFERENCE SITE) | Dataset2 | Dataset3 | Dataset4 | Dataset5 | Dataset6 |
|---|---|---|---|---|---|---|
| TOTAL | 2739 | 952 | 446 | 90 | 247 | 4402 |
| ACQUISITION PROTOCOL – FIELD STRENGTH | MPRAGE – 1.5T | MPRAGE – 3T | MPRAGE – 1.5T | SPGR – 1.5T | MPRAGE – 3T | MPRAGE – 3T |
| SCANNER | Siemens Magnetom Avanto | Philips | Siemens Avanto | GE Signa | Siemens Tim Trio | Siemens Skyra (VD13) |
| STUDY | SHIP | BLSA | AIBL | BLSA | PAC | UK Biobank |
| CITATION | [26] | [27] | [28] | [27] | [29] | [30] |

Table 3: Description of the data used for age prediction

### 4.1.2 Schizophrenia Classification

For schizophrenia classification, we use data from the PHENOM imaging consortium [31, 32], which includes scans from 3 sites, SITE 1 (n = 325), SITE 2 (n = 330), and SITE 3 (n = 180). Multiple different acquisition protocols, field strengths, and geographic locations are represented. See Table 4 for more information.

| TOTAL SUBJECTS = 825 | SITE 1 | SITE 2 | SITE 3 |
|---|---|---|---|
| TOTAL | 325 | 330 | 180 |
| CONTROLS | 170 | 173 | 105 |
| PATIENTS | 155 | 157 | 75 |
| ACQUISITION PROTOCOL – FIELD STRENGTH | MPRAGE – 3T | MPRAGE – 1.5T | BRAVO – 3T |
| CITATION | [31, 32] | [31, 32] | [31, 32] |

Table 4: Description of the data used for schizophrenia prediction

### 4.2 Preprocessing

Minimally preprocessed T1-weighted MRI scans were used in our experiments. The scans were skull-stripped using an automated multi-atlas label fusion method [33], then affinely registered to a common atlas using FMRIB's Linear Image Registration Tool FLIRT [34]. Finally, the scans underwent a quality control procedure using automatic outlier detection to flag cases for manual verification.

### 4.3 Image Harmonization to a Reference Domain

CycleGAN is a deep generative model that can be used for unsupervised image to image translation. It works by using two GANs: one to transform images from domain A to domain B, and one to transform images from domain B back to domain A. In the case of this paper, we will consider domain B to be the reference domain (the site whose variation we would like others to match) and domain A to be the external site or sites on which we want to harmonize. With CycleGAN, we learn a two-way mapping between domain A and domain B. We use the forward mapping from Domain A to Domain B to apply our harmonization and we use the reverse mapping (identity mapping) to ensure that semantic information is preserved in our original transformation. This is done by comparing our starting image in domain A to the same image that has undergone a cyclic transformation (mapped to domain B; then mapped back to domain A). We can therefore penalize the differences between the two images to ensure that any information needed to recreate the original image is preserved in the transformation to domain B. With this configuration, only information that varies on a group level, can be "memorized" by the network and discarded or changed in the transformations.

In this paper, we use a modified version (see supplemental 1) of the original Pytorch CycleGAN implementation by Jun-Yan Zhu and Taesung Park [22].

For our experiments, we use CycleGAN to learn the mapping between unpaired axial slices of scans in our training domain and an out-of-sample site that has not been seen by the prediction model. We use this mapping to harmonize scans in our new site to the training domain. In the multisite case, we modify the CycleGAN formulation by adding a conditional variable that allows the network to reconstruct the appropriate original site (See supplemental 1 for more information). This harmonized data can then be used in our prediction model.

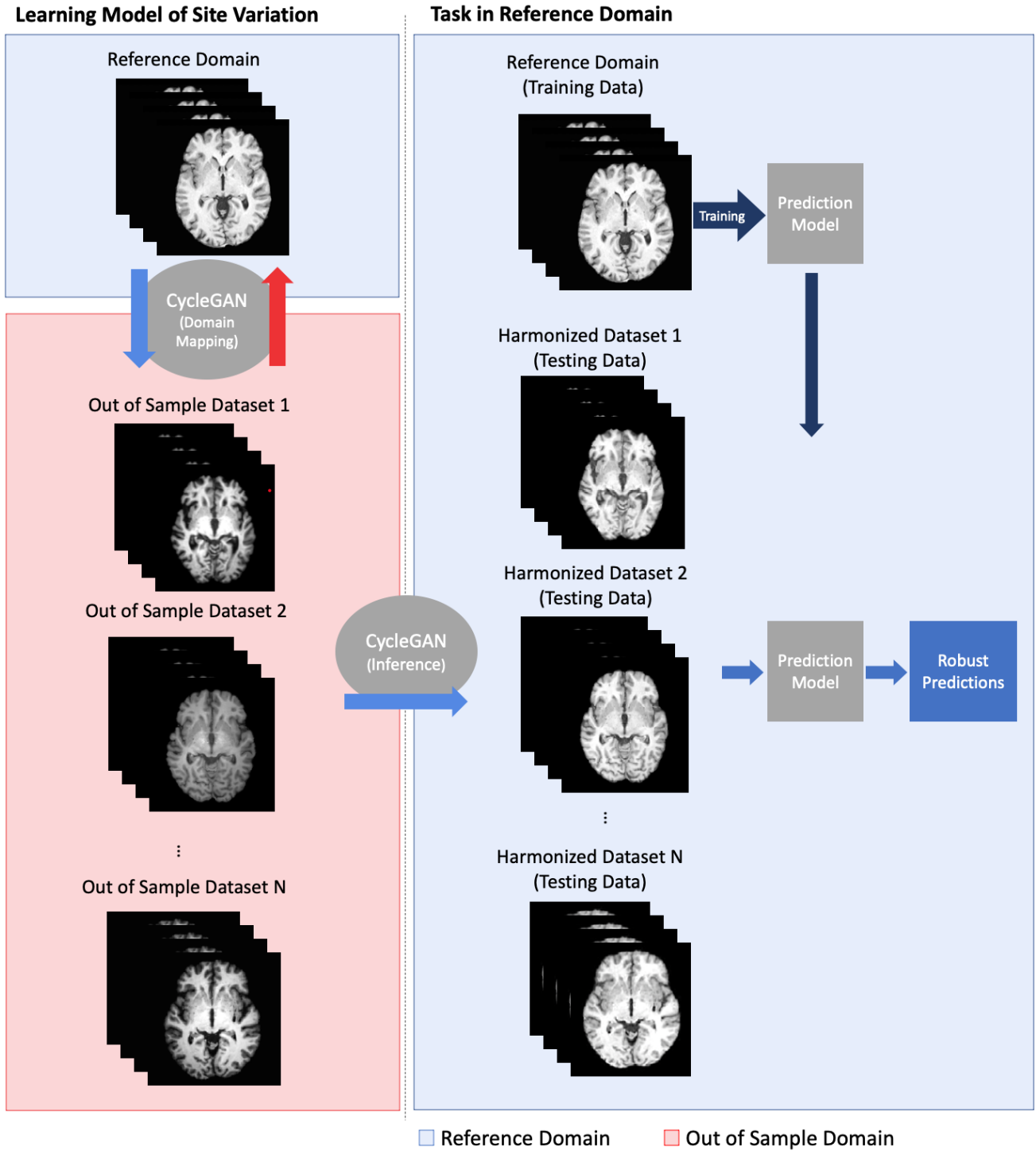

Figure 3: Description of harmonization and prediction workflow. A two-way mapping is learned between the reference domain and the out-of-sample domain or domains. The forward mapping is then applied to the out-of-sample data to obtain harmonized data. A prediction model, trained on data in the reference domain, can then be used on the harmonized data for improved generalization ability.

## 4.4 Age Prediction & Schizophrenia Classification

In the case of unharmonized experiments, the prediction model was trained only on the training site and evaluated on the test site. In the case of harmonized experiments, the prediction model was trained only on the training site and evaluated on the CycleGAN based transformation of the testing site or sites.

Both age prediction and schizophrenia classification were performed using a 10-layer ResNet model, which has been shown to perform well on a variety of imaging tasks. Our model uses a combination of convolutional and max pooling layers to incrementally reduce the dimensionality of images. After the final max pooling layer, we flatten the output and pass it through a fully connected layer of size 512 with 50% dropout and RELU activation.

In the case of age prediction, we attached a single output node with a linear activation; whereby the network can be optimized using mean squared loss. In the case of schizophrenia prediction, we attached a single output node with a sigmoid activation; whereby the network can be optimized via binary cross-entropy loss. Both models were optimized using the Adam optimizer with a learning rate of 3e-4. This learning rate is decreased by a factor of 10 when the training loss remains constant for 5 consecutive epochs. The network is considered to have converged with the training loss is constant for 10 epochs or the validation loss increases for 5 consecutive epochs.

The network is trained using the middle 80 axial slices of each MRI scan, where each slice is treated as an independent training sample. For testing, the median prediction (prediction probability for classification) of a scan is used.

## Funding

This work was supported in part by NIH grants RF1AG054409, R01EB022573, R01MH112070, R01MH120482, R01 MH113565, and NIH contract HHSN271201600059C.This work was also supported in part by the Intramural Research Program, National Institute on Aging, NIH, and the Swiss National Science foundation grant 191026. SHIP is part of the Community Medicine Research net of the University of Greifswald, Germany, which is funded by the Federal Ministry of Education and Research (grants no. 01ZZ9603, 01ZZ0103, and 01ZZ0403), the Ministry of Cultural Affairs and the Social Ministry of the Federal State of Mecklenburg-West Pomerania. MRI scans in SHIP and SHIP-TREND have been supported by a joint grant from Siemens Healthineers, Erlangen, Germany and the Federal State of Mecklenburg-West Pomerania.

## Competing Interests

HJG has received travel grants and speakers' honoraria from Fresenius Medical Care, Neuraxpharm, Servier and Janssen Cilag as well as research funding from Fresenius Medical Care.

Supplemental 1: Modifications to the CycleGAN formulation to preserve isomorphism when learning multisite data mappings.

We are constructing a function $f : \mathbb{R}^n \mapsto \mathbb{R}^n$ which maps site variation in one domain to the site variation in another, where n is the number of dimensions that our medical images vary. We define $\mathbb{R}^m \subset \mathbb{R}^n$ as the multi-dimensional space representing possible imaging variation due to site. Let $X, Y \subset \mathbb{R}^n$ where $X$ is our source domain and $Y$ is our target domain. Then $f^{-1}(Y) = \{x \in X : f(x) \in Y\}$ represents our identify function or the preimage of $f$ limited to the set $X$.

In the case where we want multiple domains to be mapped to a single template domain, we let $X_1, ..., X_k \subset \mathbb{R}^n$ be our source domains and $X_1^m, ..., X_k^m \subset \mathbb{R}^m$ represent only the site associated variation in the source domain. We want to find a function $f(X) = \{y \in Y : f(X_i) = f(X_j)$ when $X_i - X_i^m = X_j - X_j^m$ for $i, j \in 1, .., k\}$. This means that $f$ will map $X_i$ and $X_j$ to the same position when the images only differ along the $m$ dimensions of site variation. In this case our identity function $f^{-1}(Y) = \{x \in X_i : i \in 1, .., k, f(x) \in Y\}$ is a one-to-many mapping in the case where scans only differ in site related variation. Since this mapping cannot be constructed in a function we introduce a new variable $s$ that encodes site information. Our identify function then becomes $f^{-1}(Y, s) = \{x \in X_s : f(x) \in Y\}$, returning us to a bijective mapping.

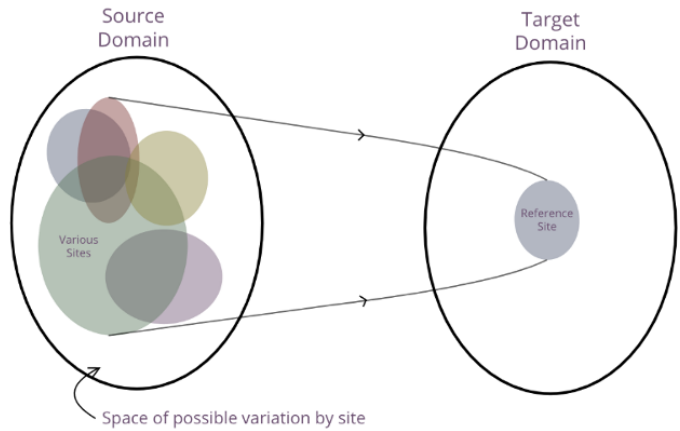

Figure 1: 2D visualization of the $\mathbb{R}^m$ space of site variation

In practice we append site information into the latent space of the identity generator. We model the site information as a learnable embedding. A vector of site information (one hot encoding) is upscaled via deconvolutions to the dimension of one channel of the latent space then appended as a new channel (prior to the transformation blocks). This restores the one-to-one mapping.